\documentclass[referee]{raa}           

\usepackage{graphicx,times}
\usepackage{natbib}
\usepackage{amssymb,amsmath}
\usepackage{enumitem}
\bibpunct{(}{)}{;}{a}{}{,}

\newcommand{\ha}{\text{H}\alpha}

\usepackage[pagebackref=true]{hyperref}

\begin{document}

   \title{Mock Observations for the CSST Mission: Multi-Channel Imager--The Cluster Field}

 \volnopage{ {\bf 20XX} Vol.\ {\bf X} No. {\bf XX}, 000--000}
   \setcounter{page}{1}

   \author{Yushan Xie\inst{1, 2},  Xiaokai Chen\inst{3}, Shuai Feng
      \inst{4}, Zhaojun Yan\inst{1}, Nan Li\inst{5}, Huanyuan Shan\inst{1,2,6}, Yin Li\inst{7}, Chengliang Wei\inst{8}, Weiwei Xu\inst{5, 9}, Zhenya Zheng\inst{1},  Ran Li \inst{9}, Wei Chen\inst{1}, Zhenlei Chen\inst{1}, Chunyan Jiang\inst{1}, Dezi Liu\inst{10}, Lin Nie\inst{1}, Xiyan Peng\inst{1}, Lei Wang\inst{8}, Maochun Wu\inst{1}, Chun Xu\inst{1}, Fangting Yuan\inst{1}, Shen Zhang\inst{5}, and Jing Zhong\inst{1}
   }

   \institute{Shanghai Astronomical Observatory, Chinese Academy of Sciences, Shanghai 200030, China; {\it hyshan@shao.ac.cn}\\
        \and
             School of Astronomy and Space Science, University of Chinese Academy of Sciences, Beijing 100049, China\\
        \and 
        Department of Astronomy, School of Physics and Astronomy, Shanghai Jiao Tong University, Shanghai 200240, China\\
        \and 
        College of Physics, Hebei Normal University, 20 South Erhuan Road, Shijiazhuang, 050024,
        China\\
        \and
        National Astronomical Observatories, Chinese Academy of Sciences, Beijing 100101, China\\
        \and
             Key Laboratory of Radio Astronomy and Technology, Chinese Academy of Sciences, A20 Datun Road, Chaoyang District, Beijing, 100101, P. R. China\\
        \and
        Department of Mathematics and Theory, Peng Cheng Laboratory, Shenzhen, Guangdong 518066, China\\
        \and 
        Purple Mountain Observatory \& Key Laboratory of Radio Astronomy, Chinese Academy of Sciences, 10 Yuanhua Road, Qixia District, Nanjing 210023, China\\
        \and
        School of Physics and Astronomy, Beijing Normal University, Beijing 100875, China\\
        \and
        South-Western Institute for Astronomy Research, Yunnan University, Kunming, Yunnan, 650500, China\\
\vs \no
   {\small Received 20XX Month Day; accepted 20XX Month Day}
}

\abstract{
The Multi-Channel Imager (MCI), one of the instruments aboard the China Survey Space Telescope (CSST), is designed to simultaneously observe the sky in three filters, covering wavelengths from the near-ultraviolet (NUV) to the near-infrared (NIR). With its large field of view ($7.5^{\prime}\times7.5^{\prime}$), MCI is particularly well-suited for observing galaxy clusters, providing a powerful tool for investigating galaxy evolution, dark matter and dark energy through gravitational lensing. Here we present a comprehensive simulation framework of a strong lensing cluster as observed by MCI, aiming to fully exploit its capabilities in capturing lensing features. The framework simulates a strong lensing cluster from the CosmoDC2 catalog, calculating the gravitational potential and performing ray-tracing to derive the true positions, shapes and light distribution of galaxies within the cluster field. Additionally, the simulation incorporates intra-cluster light (ICL) and spectral energy distributions (SEDs), enabling further strong lensing analyses, such as ICL seperation from galaxy light and mass reconstruction combining strong and weak lensing measurements. This framework provides a critical benchmark for testing the MCI data pipeline and maximizing its potential in galaxy cluster research.
\keywords{software: simulations --- software: public release --- methods: numerical ---  surveys --- virtual observatory tools 
}
}

   \authorrunning{Y.-S. Xie et al. }            
   \titlerunning{An Overview of CSST-MCI Cluster Field Simulation}  
   \maketitle

%
\section{Introduction}           
\label{sect:intro}
The China Survey Space Telescope \citep[CSST, ][]{zhan21, zhan11, cao18, gong19} is a space telescope orbiting the Chinese Space Station (CSS). Aiming at addressing fundamental questions in cosmology and astrophysics, such as the large-scale structure of the Universe and the mysteries of dark matter and dark energy, CSST will be equipped with several modules, including the main Survey Camera (SC), a Multi-Channel Imager (MCI), an Integrated Field Unit spectrometer (IFS), an exoplanet imaging coronagraph, and a Tera-hertz module. 

Among these modules, the MCI offers simultaneous imaging across three filters (CBU: $0.255 - 0.43\mu \rm m$, CBV: $0.43 - 0.7 \mu \rm m$, CBI: $0.7 - 1.0 \mu \rm m$) that cover near-ultraviolet (NUV), optical and near-infrared (NIR) wavelengths, and a large field-of-view (FoV) of $7.5^{\prime} \times 7.5^{\prime}$ with a pixel size of $0.05^{\prime\prime}$. These design features make MCI particularly effective for capturing high-resolution, multi-wavelength images of strong lensing clusters, facilitating a variety of detailed studies with the observed multiple images, which provide valuable insights into the Universe, e.g., mapping the dark matter distribution within clusters \citep{massey15, jauzac19, bergamini23, meneghetti20, meneghetti23, Kawai24}, investigating the nature and behavior of dark energy \citep{jullo10, magana18, caminha21}, studying the formation and evolution of galaxies \citep{welch23, asada24, mowla24}, and exploring the expansion rate of the Universe \citep{grillo18, acebron23, pascale24, suyu17, courbin18, millon20}. 

Interpreting strong lensing features requires careful lens modeling based on a comprehensive understanding of astrophysical and instrumental systematic errors \citep{acebron17, chirivi18, granata22}. In this work, we create a simulated cluster field and employ a ray-tracing algorithm \citep{li16} to model the lensing effects for all line-of-sight galaxies within the field. This simulation serves as a testbed for validating the MCI data pipeline by incorporating detailed instrumental effects, which is discussed in a companion paper (Yan et al., hereafter Yan25), and for evaluating the performance of various lens modeling algorithms, e.g., \textsc{Lenstool} \citep{jullo07}, \textsc{CURLING} \citep{xie24}, while also identifying the impact of other sources of systematics. As the contamination from ICL \citep{Zwicky1937, DeMaio2015, Montes2019}, which consists of diffuse light emitted by stars unbound to any individual cluster galaxy,  poses a significant challenge in extracting the light of lensed arcs and member galaxies \citep{montes14, dicris17} -- both critical components for accurate lens modeling, we include ICL in our simulation to enable us to develop strategies for disentangling these light components in real observations. Furthermore, MCI’s large FoV allows observations of cluster outskirts extending to several Mpc, providing valuable weak lensing information. By combining weakly and strongly lensed galaxies, we can achieve a more comprehensive mass reconstruction of the cluster \citep{quinn18, strait18, niemiec23, patel24}. Since photometry is essential for both measurements -- particularly for weak lensing -- we incorporate SEDs for all galaxies, ensuring that photometric measurements are available for further weak lensing analyses with the simulated cluster. 

The work is structured as follows. In Section~\ref{sect2} we introduce the galaxy data from the CosmoDC2 catalog, which serves as the foundation for our simulation framework. In Section~\ref{sect3}, we introduce the methodology for constructing mock images of the simulated cluster within the framework, outlining the steps involved in incorporating gravitational lensing effects. Section~\ref{sect4} expands the simulation framework by introducing additional components, such as the cluster ICL, the SEDs of the galaxies, and simplified instrumental effects. Finally, Section~\ref{sect5} provides a summary and conclusion. Throughout the paper, we assume a flat $w$CDM cosmology ($\Omega_{\Lambda,0} = 0.7$, $\Omega_{\rm m}=0.3$, $h$ = 0.7) with a constant dark energy equation of state parameter $w = w_{0} = -1$.

\begin{figure*}
    \centering
    \includegraphics[width=\textwidth]{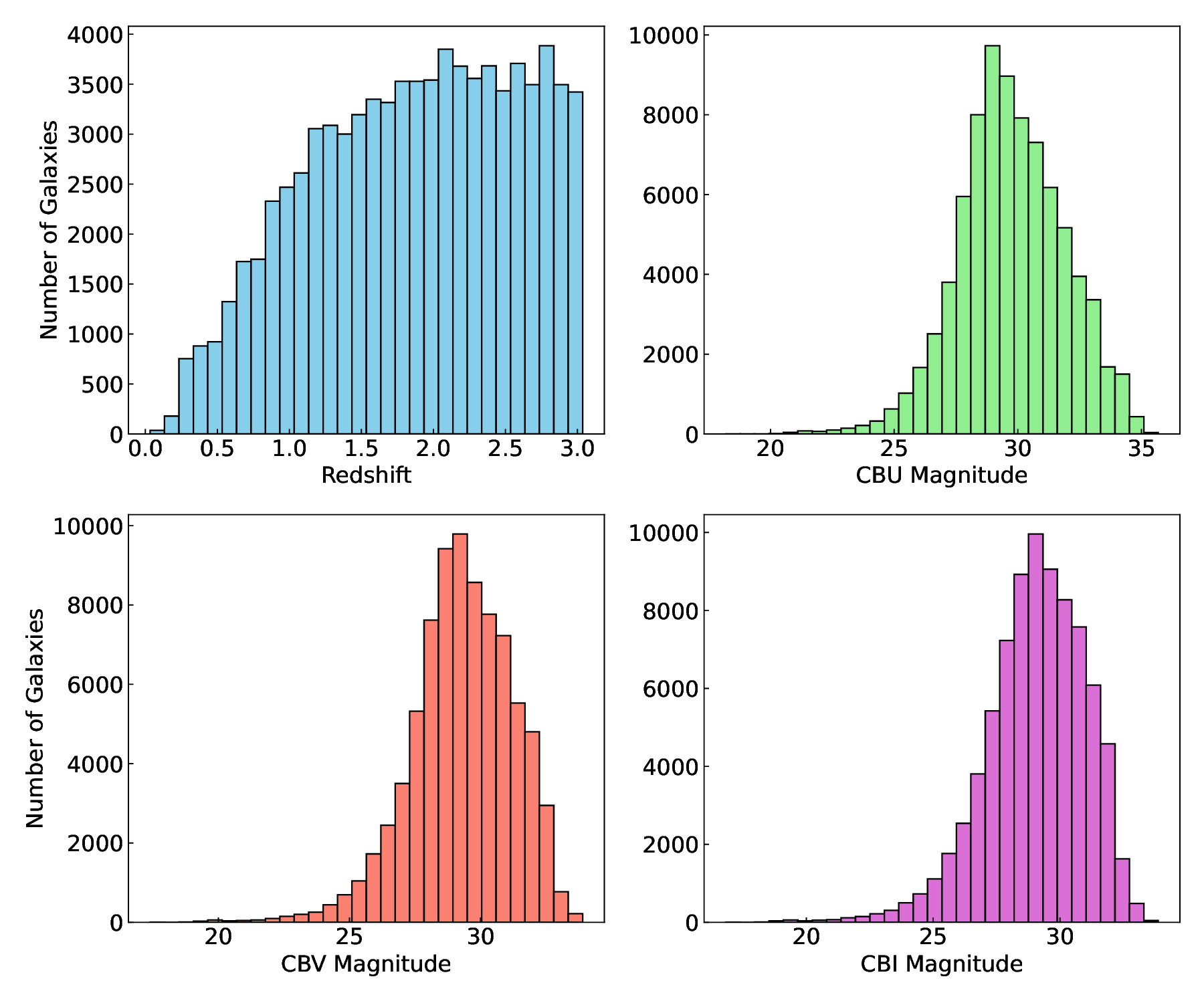}
    \caption{Redshift and CBU, CBV, CBI magnitude distributions of the 80,532 source galaxies within the $7.5^{\prime}\times7.5^{\prime}$ lightcone of the simulated MCI cluster.}
    \label{fig1}
\end{figure*}

\section{Cluster Catalog and Simulation Setup} \label{sect2}

In this work, the simulated cluster field for MCI is constructed based on the CosmoDC2 \citep{cosmodc2} catalog, which provides information about the cluster member galaxies and the line-of-sight galaxies.

\subsection{Cosmological Simulation}

The CosmoDC2 catalog is created from the `Outer Rim’, a large volume cosmological N-body simulation with a box size of 4.225 Gpc and $(10,240)^3$ tracer particles, achieving a mass resolution of $m_{p}=2.6\times10^9M_{\odot}$. 

Halo catalogs are generated using a parallel Friends-of-Friends (FoF) halo finder with a linking length of $b = 0.168$ and a minimum requirement of 20 particles per halo. These halos are then linked across snapshots using a particle-membership algorithm \citep{rangel20} to construct the halo merger tree, which serves as the foundation for the construction of halo lightcones. Galaxies are placed within these halo lightcones by first resampling from the galaxy catalog produced by the \textsc{UniverseMachine} \citep{universemachine} code and refining them through empirical modeling to ensure realistic property distributions. Each galaxy is then matched with a counterpart in the galaxy library generated with the \textsc{Galacticus} \citep{galacticus} code to incorporate a comprehensive set of physical properties. 

The final catalog contains $\sim$ 2.26 billion galaxies in a 440 $\rm deg^2$ field that spans $0 < z < 3$. Each galaxy has 551 listed properties including stellar mass, fluxes in LSST and SDSS bands, shape information, etc.

\subsection{Galaxies in the CosmoDC2 Catalog}
\subsubsection{Cluster Member Galaxies} \label{sect221}

The MCI cluster simulation is performed by extracting the cluster galaxies from the CosmoDC2 catalog. We focus on a cluster with a virial mass of $\rm M_{vir} = 1.13 \times 10^{15} M_{\odot}$ at redshift of $z_{\rm L}=0.3$, which stands for a typical cluster for strong gravitational lensing studies. The cluster field is cropped to a region of $7.5^{\prime}\times7.5^{\prime}$ to align with the MCI observational setup. Within this field, we identify a total of 211 cluster member galaxies. Detailed galaxy properties from the catalog include positions, morphological parameters (semi-major and semi-minor axes, position angle, S$\rm \acute{e}$rsic profile index), and magnitudes in the LSST $g$, $r$, and $i$ bands, which represent the corresponding MCI CBU, CBV, and CBI bands. The morphological and flux parameters are further decomposed for disk, bulge, and total components, with each galaxy also assigned a bulge-to-disk ratio. These data are used to compute the gravitational potential of the cluster for ray tracing and to generate the MCI-observed image of the cluster field in the subsequent analyses.

\subsubsection{Catalog of Source Galaxies} \label{sect222}

Within the $7.5^{\prime}\times7.5^{\prime}$ lightcone of the simulated cluster, a total of 80,532 source galaxies are identified. Each of these galaxies (we classify them as `line-of-sight galaxies' in the following analyses) is characterized by the same set of properties as the cluster member galaxies described in Section~\ref{sect221}. In addition, the redshift of each source galaxy is recorded to facilitate the computation of its light distribution after being gravitationally lensed by the cluster potential. The source galaxies span a redshift range from $z_{\rm s}=0$ to $z_{\rm s}=3$. To illustrate their overall distribution, Fig.~\ref{fig1} presents the redshift and magnitude distributions across the MCI CBU, CBV, and CBI bands for all line-of-sight galaxies.

\section{Lensing Simulations Framework} \label{sect3}
\subsection{Lensing Theory} 

Gravitational lensing \citep{kneib11, meneghetti21} is a powerful astrophysical phenomenon that occurs when the light from a distant source, such as a galaxy or a quasar, is deflected by the gravitational field of a massive object, such as a galaxy cluster, lying between the source and the observer. This effect, predicted by Einstein's General Theory of Relativity, can magnify, distort the source, and create multiple images of the background sources. 

To accurately account for the lensing effect, we perform ray-tracing, a method that traces the path of light rays as they travel through the gravitational field of the cluster. The process involves solving the lens equation, which relates the positions of the source $\beta$ and the image $\theta$,  
\begin{equation}\label{eq3}
    \vec{\beta} = \vec{\theta} - \vec{\alpha}(\vec{\theta}),
\end{equation}
where the deflection angle $\vec{\alpha}(\vec{\theta})$ is determined by the Newtonian gravitational potential $\Phi(\vec{\theta})$ and the geometry of the Universe, specifically the angular diamater distances between the lens $L$, source $S
$, and observer $O$ ($D_{LS}$, $D_{OL}$, and $D_{OS}$), 
\begin{equation}\label{eq4}
    \vec{\alpha}(\vec{\theta}) = \frac{2}{c^2}\frac{D_{LS}}{D_{OL}D_{OS}} \nabla \Phi(\vec{\theta}).
\end{equation}
The lens equation is typically highly non-linear and can have multiple solutions, corresponding to the cases of multiple images as a result of strong gravitational lensing.

In addtion to multiple images, another important feature of lensing distortion is the magnification of the sources. Mapping the coordinates from the source plane to the image plane involves the \textit{Jacobian Matrix},
\begin{equation}
    A \equiv \frac{\partial \vec{\beta}}{\partial \vec{\theta}} = (\delta_{ij} - \frac{\partial \alpha_{i}(\vec{\theta})}{\partial \theta_{j}}) = (\delta_{ij} - \frac{\partial ^ 2 \Psi (\vec \theta)}{\partial \theta_{i} \partial_{j}})= \begin{pmatrix}
1 - \kappa - \gamma_{1} & - \gamma_{2}\\\\
    -\gamma_{2} & 1 - \kappa + \gamma_{1}
\end{pmatrix}, 
\end{equation}
where the lensing potential $\Psi(\boldsymbol{\theta}) = \frac{2}{c^2} \frac{D_{LS}}{D_{OL} D_{OS}} \int \Phi(D_l \boldsymbol{\theta}, z) \, dz,
$, is derived from the gravitaional potential $\Phi$ by integrating along the line of sight, the convergence term $\kappa = \frac{1}{2}\Delta \Psi = \frac{1}{2}(\Psi_{11} + \Psi_{22})$ defines the isotropic transformation of images, and the shear terms $\gamma_1 = \frac{1}{2}(\Psi_{11} - \Psi_{22})$ and $\gamma_2 = \Psi_{12} = \Psi_{21}$, with the complex shear $\gamma = \gamma_{1} + i\gamma_{2}$, delineate the stretching of images. The magnification is given by the inverse of the determinant of the Jacobian matrix,
\begin{equation}
    \mu \equiv {\rm det} M = \frac{1}{{\rm det} A} = \frac{1}{(1-\kappa)^2 - |\gamma|^2}, 
\end{equation}

\begin{figure*}
    \centering
     \includegraphics[width=\textwidth]{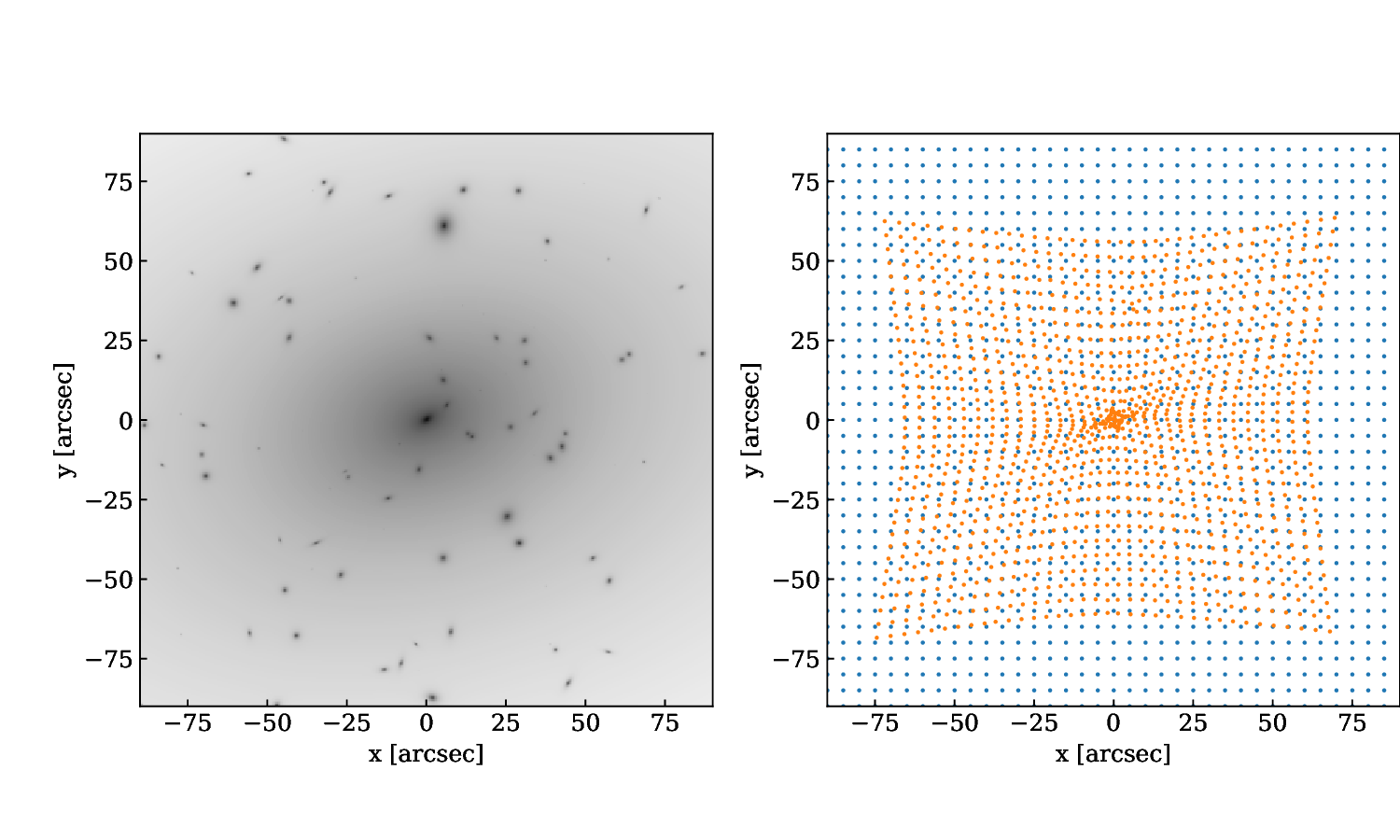}
    \caption{Left panel: Convergence map of the simulated MCI cluster. Right panel: Illustration of the light rays on the lens plane (blue dots) and their mapped positions on the source plane (orange dots) due to the deflection of gravitational lensing.}
    \label{fig-convergence}
\end{figure*}

\subsection{Cluster Potential}

According to Equations~\ref{eq3} and \ref{eq4}, the solution of the lens equation for ray tracing is based on computing the lensing potential. In this simulation, the cluster is a combination of several components, each contributing to the overall gravitational potential:
\begin{itemize}
    \item Dark matter halo: The dominant component of the cluster’s mass distribution is the dark matter halo, which is associated with the brightest cluster galaxy (BCG). We model this halo using the Navarro-Frenk-White (NFW) profile \citep{Navarro1997},
    \begin{equation}
         \rho(r) = \frac{\rho_s}{\frac{r}{r_s} \left( 1 + \frac{r}{r_s} \right)^2},
    \end{equation}
    where $\rho_{s}$ is the characteristic density of the profile, $r_{s} = r_{200} / c$ is the characteristic radius, with $r_{200}$ defined as the radius within which the average density of the cluster is 200 times the critical density of the Universe ($\rho_{c}$) at the cluster redshift, $c$ defined as the concentration parameter.
     As provided in the catalog, the halo has a virial mass of $\rm M_{vir}=1.13 \times 10^{15} M_{\odot}$, and a concentration of $c=3.86$.
    \item Subhalos: We employ the Pseudo-Isothermal Elliptical Mass Distribution \citep[PIEMD,][]{eliasdottir07} to model the subhalos associated with member galaxies. The PIEMD potential is defined as,
    \begin{equation}
	\rho(r) = \frac{\rho_{0}}{(1+\frac{r^2}{r_{\rm core}^2})(1+\frac{r^2}{r_{\rm cut}^2})}, 
    \end{equation}
    where $\rho_{0}$ denotes the central density of the profile, while $r_{\rm core}$ and $r_{\rm cut}$ are the core radius and the truncation radius, respectively. Within the regions where $r_{\rm core} < r < r_{\rm cut}$, the profile behaves as $\rho \sim r^{-2}$, while it transitions to $\rho \sim r^{-4}$ in the outer regions where $r > r_{\rm cut}$. Given that PIEMD parameters, i.e., $\rho_0$, $r_{\rm core}$, and $r_{\rm cut}$, are not provided in the original CosmoDC2 catalog, we assign these values to each galaxy based on its luminosity, using the scaling relations \citep[see Equation~4 in][]{jullo07} as used in strong lensing modeling:
    \begin{align}\label{eq2}
    \begin{split}
        \left \{
        \begin{array}{ll}
        \rm  \sigma^{gal}_{\it{i}}=\sigma^{ref} 10^{0.4 \frac{m^{ref}-m_{\it{i}}}{\alpha}}\\
        \rm \it{r}^{\rm gal}_{core,\it{i}}=r\rm _{core}^{ref}10^{0.4 \frac{m^{ref}-m_{\it{i}}}{2}}\\
        r\rm ^{gal}_{cut,\it{i}}=\it{r}\rm _{cut}^{ref}10^{0.4 \frac{2(m^{ref}-m_{\it{i}})}{\beta}},
        \end{array}
        \right.
    \end{split}
    \end{align}
    The typical core radius, cut radius and velocity dispersion for the scaling relations are set to $r\rm _{core}^{ref} = 1.0\ kpc$, $r\rm _{cut}^{ref} = 12.0\ kpc$, $\sigma^{\rm ref} = 200.0\ \rm km/s$ in this work.
    \end{itemize} 
The deflection angle map, calculated as the gradient of the lensing potential (Equation~\ref{eq4}), is then used to derive the image distortions for each source in the catalog (Equation~\ref{eq3}). In Figure~\ref{fig-convergence}, we show the convergence map of the simulated cluster and the map of light rays from the lens plane to the source plane, illustrating the effects of gravitational lensing.

\begin{figure*}
    \centering
    \includegraphics[width=\textwidth]{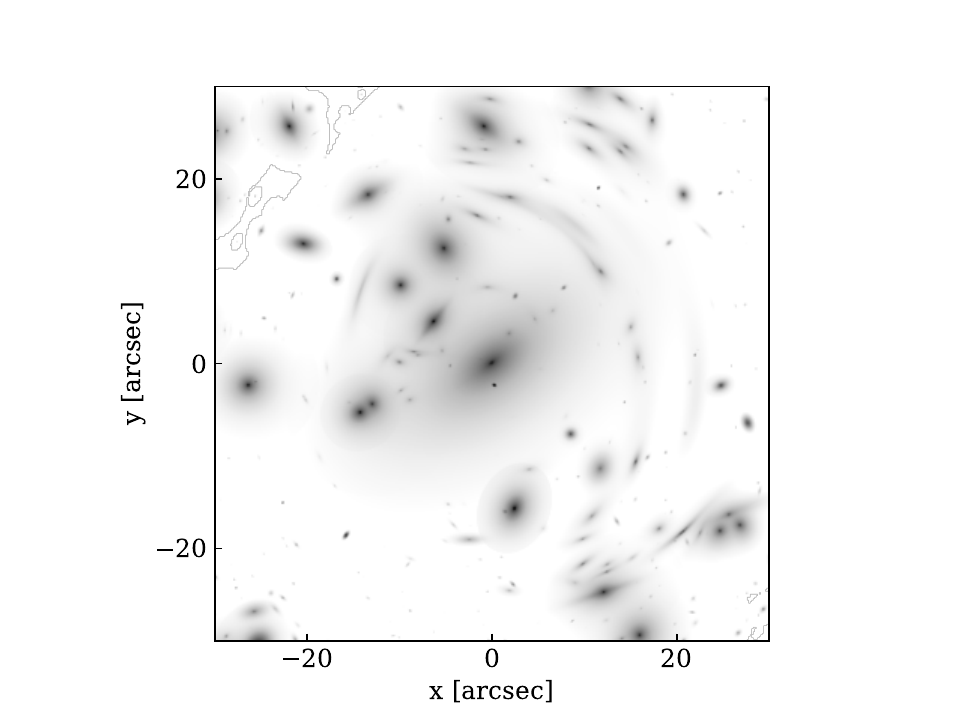}
    \caption{Galaxy distributions within the central region ($60^{\prime\prime} \times 60^{\prime\prime}$) of the simulated cluster, accounting for the gravitational lensing effect.}
    \label{fig-lens}
\end{figure*}

\subsection{Ray Tracing for Line-of-sight Galaxies}

To generate the mock image, we consider the light distribution of every galaxy within the field. Each galaxy is modeled as a combination of a bulge and a disk, both described by a S$\rm \acute{e}$rsic profile: 
\begin{equation}
	I(r) = I_e \exp\left\{ -b_n \left[ \left( \frac{r}{r_e} \right)^{1/n} - 1 \right] \right\},
\end{equation}
where $I(r)$ is the intensity at radius $r$, $I_{e}$ is the central intensity, $r_{e}$ is the effective radius, $n$ is the S$\rm \acute{e}$rsic profile index, and $b_{n}$ is a constant parameter. As detailed in Section~\ref{sect222}, the S$\rm \acute{e}$rsic profile parameters, along with the galaxy’s position, shape, and magnitude, are provided by the CosmoDC2 catalog. 

For galaxies lying behind the cluster, we calculate the observed source position $\theta$ from its original position $\beta$ based on the deflection angle $\alpha$, accounting for the gravitational lensing effect. Here, we assume the lens as the single-plane cluster, as the primary goal is to compare lens modeling codes and support companion work that focuses on MCI instrumental simulation (Yan25). More complex scenarios involving multi-lens planes will be explored in future studies. 

Once the light distributions of all galaxies have been computed, we superimpose them to generate the complete simulated image of the cluster field. In Figure~\ref{fig-lens}, we present the generated image in the MCI CBV band. For clarity, only the central $60^{\prime \prime} \times 60^{\prime\prime}$ is displayed, highlighting the lensing features.

\begin{figure*}
    \centering
    \includegraphics[width=\textwidth]{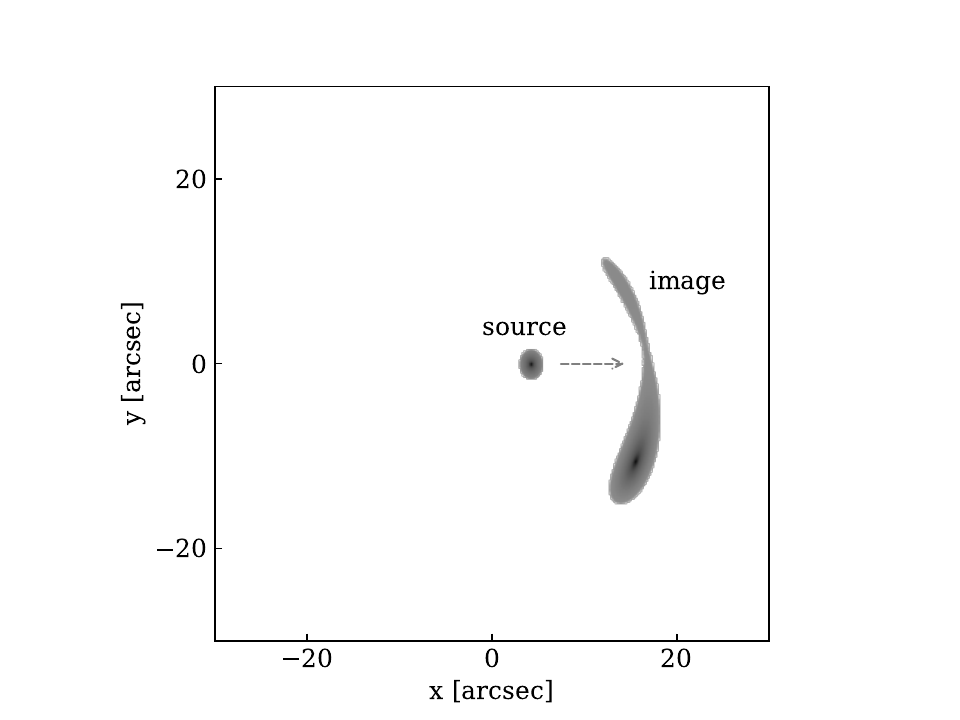}
    \caption{An illustration of the multiply lensed arc. The `source' shows where the source is originally lying. The `image' indicates its appearance and position after being deflected by the lens.}
    \label{fig-arc}
\end{figure*}

In addition to ray tracing the galaxies provided by the CosmoDC2 catalog, we further add twenty galaxies to incorporate strongly lensed arcs, which will be valuable for strong lensing analyses with this simulated cluster. These galaxies are randomly extracted from the same catalog, while we adjust their positions to guarantee the formation of giant arcs. An example of a source galaxy and its corrsponding lensed arc is shown in Figure~\ref{fig-arc}. Note that in real observations, the source and the lensed image cannot be seen simultaneously. Here, they are presented together for clarity in illustrating positional and morphological changes.

\subsection{Computational Accelerating with JAX}

Given that the catalog within the MCI FoV contains over 80,000 sources, and the lensing potential is computed across a grid of $9,000 \times 9,000$ pixels, the process of generating the simulated images originally requires over 10 hours, which is highly time consuming. To address this, we optimize the simulation algorithm based on the JAX framework \citep{jax2018github}. This enhancement allows us to perform ray-tracing for an entire simulated cluster system in just 6 minutes, achieving a two-order-of-magnitude speedup compared to the previous implementation.

\section{Simuation Enhancements} \label{sect4}

To prepare for MCI observations of galaxy clusters, we simulate a cluster with strong lensing effect, as introduced in the previous sections. This simulated cluster serves as an essential testbed for the MCI team to implement realistic instrumental effects, such as point-spread function (PSF) convolution, charge transfer inefficiency (CTI),  nonlinearity, and other detector effects, which are detailed in the companion paper by Yan25. Additionally, simulated multiple-image families, coupled with the known cluster lens potential, provide a valuable resource for testing and comparing the performance of various lens modeling codes, which can help with the development of modeling algorithms in the strong lensing community. Beyond the primary simulated cluster for scientific validation, we also incorporate several additional simulations to facilitate further testing with the cluster in the context of MCI. These additional simulations cover aspects including the ICL of the cluster, and the SEDs for all member galaxies and line-of-sight galaxies, and the implementation of instrumental effects. 

\begin{figure*}
    \centering
    \includegraphics[width=\textwidth]{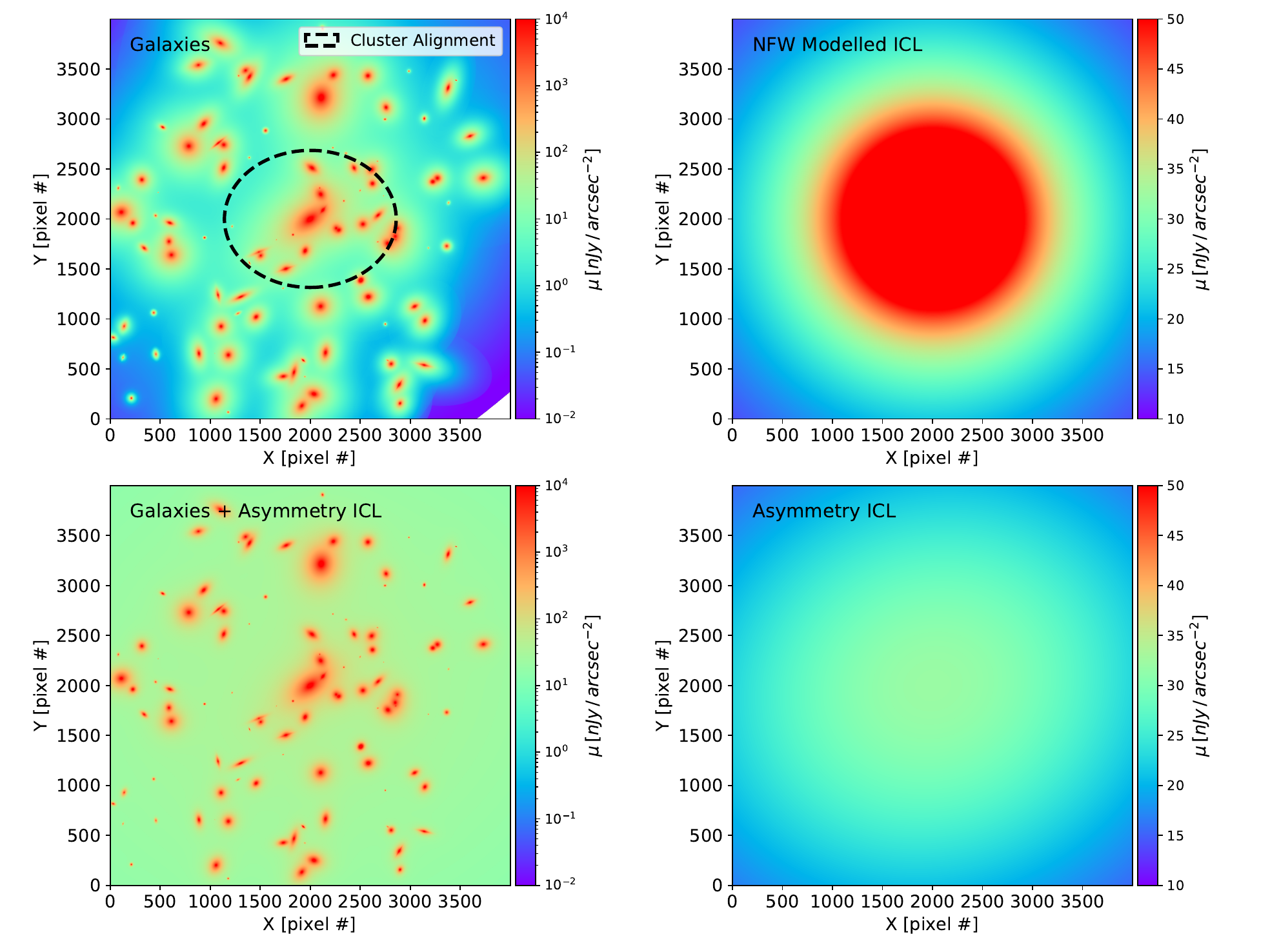}
    \caption{Illustration of the ICL mock in the CBV band. The image is presented within $4000 {\times} 4000$ pixels (corresponding to $200^{\prime\prime} {\times} 200^{\prime\prime}$). Top Left: Luminosity distribution of cluster member galaxies. The second-order moment of the galaxy position distribution is represented by the black dashed ellipse in the center of the figure. Top Right: Simulated ICL image modeled using the NFW profile. Bottom Left: Mock of asymmetric ICL luminosity distribution. Bottom Right: The luminosity distribution of the sum of the asymmetric ICL and the member galaxies of the cluster.}
    \label{fig-icl}
\end{figure*}

\subsection{Simulation of Intra-Cluster Light}

ICL is the diffuse and faint stellar population that spreads through clusters of galaxies, stripped from their host galaxies. First observed by Zwicky in 1937 \citep{Zwicky1937} and later confirmed in the Coma Cluster \citep{Zwicky1951, Zwicky1957}, ICL has become an essential feature in the study of galaxy clusters. Over the last two decades, numerous studies have shown that the ICL is closely related to the nature of galaxy clusters, in particular their dynamical evolution history \citep{DeMaio2015, DeMaio2018, Montes2019, Contini2014, Contini2018, Contini2023, Chen2022}. The light of ICL is often entangled with the light from galaxies, as well as with lensed arcs \citep{montes14, dicris17}, complicating the extraction of these features for strong lensing analyses. Moreover, the ICL can also affect the estimation of galaxy shapes and counts in weak lensing regime \citep{Gruen2019}. Facing this challenge, we incorporate the ICL into our simulated cluster field, aiming to refine algorithms for accurately subtracting its contribution.

To simulate the ICL, we first model both the mass and luminosity distributions of this diffuse stellar population. The mass distribution of the ICL is assumed to follow the NFW profile \citep{zibetti05, pillepich18, zhang19}, with the parameters $M_{\rm ICL}$ and $c_{\rm ICL}$. The total mass of the ICL is estimated using an empirical relation from Equation~(5) in \cite{Purcell2007}, which yields the ICL fraction $f_{\rm ICL} = M_{*}^{\rm diffuse} / M_{*}^{\rm total}$, with $M_{*}^{\rm total}$ defined as the total stellar mass of the cluster, and $M_{*}^{\rm diffuse}$ defined as the mass of the ICL. We calculate $f_{\rm ICL} = 0.2162$, $M_{*}^{\rm total}= 2.16 \times 10^{13} M_{\odot}$, $M_{*}^{\rm diffuse}= 4.66 \times 10^{12} M_{\odot}$, and make a simpilified assumption that the ICL follows the concentration of the dark matter halo, with $c = 3.86$, although studies suggest that the ICL is typically more concentrated than the total mass distribution \citep{wang21, rogo16, lowing15, cooper13}. 
After constructing the mass distribution, we simulate the luminosity distribution of the ICL, for the stellar mass-light ratio, we follow the coefficient in the modified relation of Equation~2 in \cite{Chen2022}, which approximately equals to unity. We then account for the effects of cosmic dimming and k-corrections \citep{hogg99},
\begin{equation}
    \mu = \mu_{0} \left( \frac{M_{h} }{ M_{h,0} } \right)^{-1 / 3} \times \left( \frac{D_{a} }{D_{a,0} } \right)^{2} \times \left( \frac{D_{l,0} }{D_{l} } \right)^{2} \times 10^{-0.4(m - m_{0}) }
\end{equation}
where $\mu_{0}$ is the surface brightness profie of a cluster at redshift $z_{0}$ ($D_{a,0}$ and $D_{l,0}$ are the corresponding angular diameter distance and luminosity distance, respectively), with mass $M_{h,0}$ and rest-frame magnitude $m_{0}$. The term $\left( \frac{D_{a} }{D_{a,0} } \right)^{2} \times \left( \frac{D_{l,0} }{D_{l} } \right)^{2} $ accounts for the cosmic dimming effect, while $10^{-0.4(m - m_{0}) }$ represents the k-correction term.

Initially, the ICL distribution is assumed to be symmetric. To simulate a more realistic scenario, we introduce asymmetries into the ICL distribution based on the spatial distribution of cluster galaxies. This is achieved by computing the second-order moment of the spatial distribution of the member galaxies and generating a normalized two-dimensional Gaussian function with these parameters. This Gaussian is then convolved with the symmetric ICL light distribution, incorporating the alignment with the observed dispersion of cluster galaxies.

Figure~\ref{fig-icl} illustrates the ICL simulation, showing the symmetric and asymmetric light distributions (upper right and lower left panels), along with the superimposed ICL and galaxy distributions (lower right panel). 

\begin{figure*}
    \centering
    \includegraphics[width=0.98\linewidth]{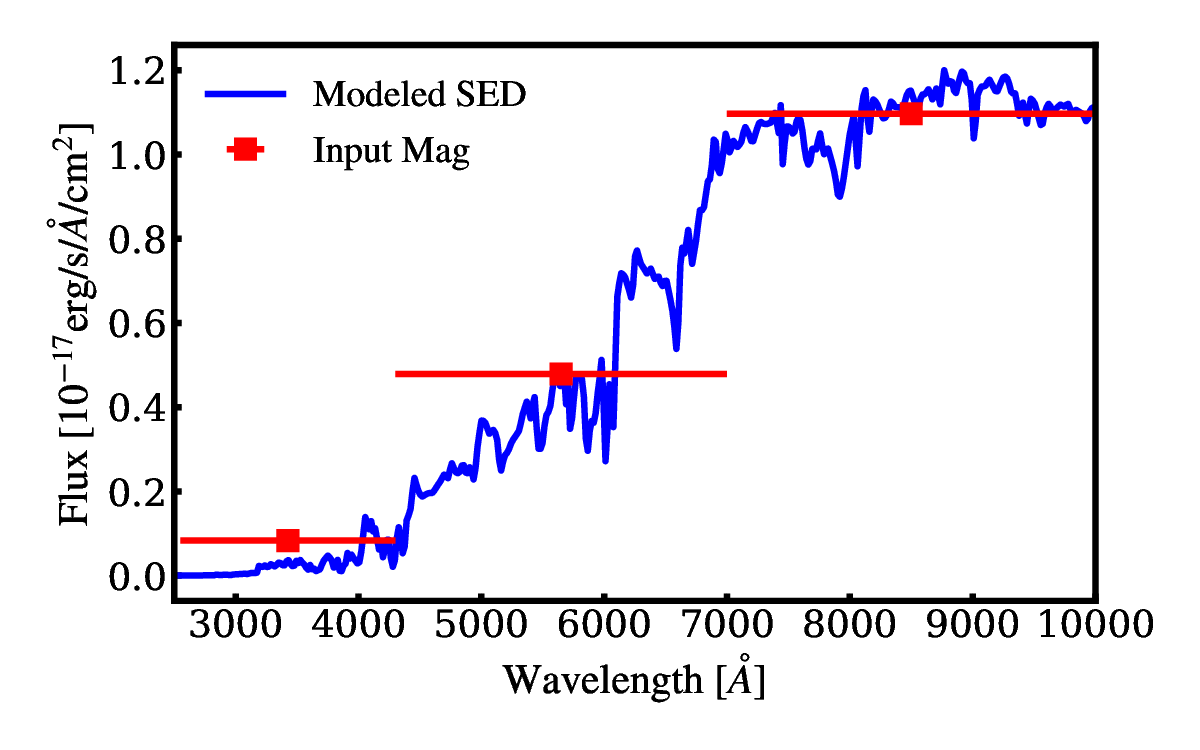}
    \caption{An example of SED mocking. The three red squares represent the input magnitudes ($m_\text{CBU} = 24.1$, $m_\text{CUV} = 22.2$, and $m_\text{CBI}=21.3$) of a galaxy with a redshift of $z = 0.53$. The error bars represent the wavelength coverage of three bands. The blue line is the modeled spectrum of this galaxy.}
    \label{fig:case_sed}
\end{figure*}

\subsection{Spectral Energy Distributions for Line-of-sight Galaxies}

The MCI's notable advantages over other space-based observations include its designed depth up to 30 AB magnitude in optical and NUV bands, along with the large FoV of $7.5^{\prime} \times 7.5^{\prime}$. These characteristics enable a comprehensive mass map of galaxy clusters, extending several Mpc, by complementing weak lensing measurements with strong lensing observations. Consequently, accurate photometry for the large amount of weakly lensed sources is necessary to interpret their properties. Here, we present the simulation of SEDs for all the line-of-sight galaxies within the cluster field, to take full advantage of MCI's multi-band photometric capabilities, particularly its narrow-band sensitivity. Since the galaxies provided in the original CosmoDC2 catalog are generated by combining the \textsc{UniverseMachine} and \textsc{Galacticus} models, and their star formation histories (SFHs) are not necessarily consistent or complete, we generate the SEDs based on the photometries.

We begin by constructing SED templates based on single stellar population (SSP) spectra from \citep{Maraston2005}. The SSP templates span wavelengths from $100\AA$ to $10,000\AA$, with stellar population ages ranging from $10^3$ yr to 15 Gyr, and metallicities from [Z/H] = -2.25 to [Z/H] = 0.67. In this study, only four values of metallicity, [Z/H] = (-1.35, -0.33, 0, 0.35), are considered, which are sufficient to simulate the SEDs for the vast majority of galaxies. For a given metallicity, we stack the SSP into a composite stellar population (CSP) spectrum according to an exponentially declining star formation history, 
\begin{equation}
    \text{SFR}(t) \sim e^{-(t - t_0)/\tau}
\end{equation}
where $\tau$ is fixed at $1.0$ Gyr, and $t_0$ is set as $8$ typical values from $10^7$ yr to $12$ Gyr. The fluxes in the CSP are set to zero for rest wavelengths shorter than $912\AA$ to model the absorption of ultraviolet photons by cold gas within galaxies. This process results in 32 continuous spectra of the stellar population, covering 4 metallicities and 8 stellar ages.

Next, we used Gaussian functions to construct the emission line templates for ionized gas within galaxies. The line widths of the emission lines are uniformly taken to correspond to a velocity dispersion of $\rm 250 \ km/s$, whereas the flux ratios of the emission lines are determined by the metallicity of the galaxies (for details, refer to Feng et al. in preparation). To match the stellar continuum templates, the ionized gas templates also include four metallicities. The final SED template is constructed by superimposing the emission line spectrum with the corresponding metallicity on the stellar population continuum spectrum. During the superimposition, the intensity of the emission line spectrum is determined by the equivalent width of $\ha$, $\text{EW}(\ha)$, which can be approximated by \citep{Fumagalli2012,Marmol2016}:
\begin{equation}
    \text{EW}(\ha)/\AA = 100 (t_0 / \text{Gyr}) ^ {-1.5},
\end{equation}
where $t_{0}$ is the stellar population age of the galaxy.

Using the SED templates, we simulate the SED of a galaxy in the wavelength range from $2,500\AA$ to $10,000\AA$ according to the apparent magnitudes in the three MCI bands and the redshift of the galaxy. The overall shape of the SED is first simulated. For a galaxy with a redshift of $z$, we need to shift all $32$ SED templates at the rest wavelength to the redshift $z$ and calculate the magnitudes of the redshifted SED templates in the three bands of CBU, CBV, and CBI. The best SED template is selected which minimizes the difference between the galaxy and SED template magnitudes. Secondly, we consider the change in the shape of the SED caused by dust attenuation. We compare the $\text{CBU}-\text{CBI}$ color of the best template with that of the input catalog. If the $\text{CBU}-\text{CBI}$ color in the input catalog is redder, we redden the best template according to the extinction law of \citet{Calzetti1994} to obtain the observed SED shape of the target source. If not, the best template will be adopted as the observed SED shape. Finally, we calibrate the absolute flux of the best template using the magnitude in the CBV band. In Fig.~\ref{fig:case_sed}, we display one example of the SED simulation. The three red squares represent the apparent magnitudes in the MCI bands provided by the input catalog, the blue spectrum shows the simulated complete SED of the galaxy, demonstrating the fit between the data and the simulated SED template.

\begin{figure*}
    \centering
    \includegraphics[width=\textwidth]{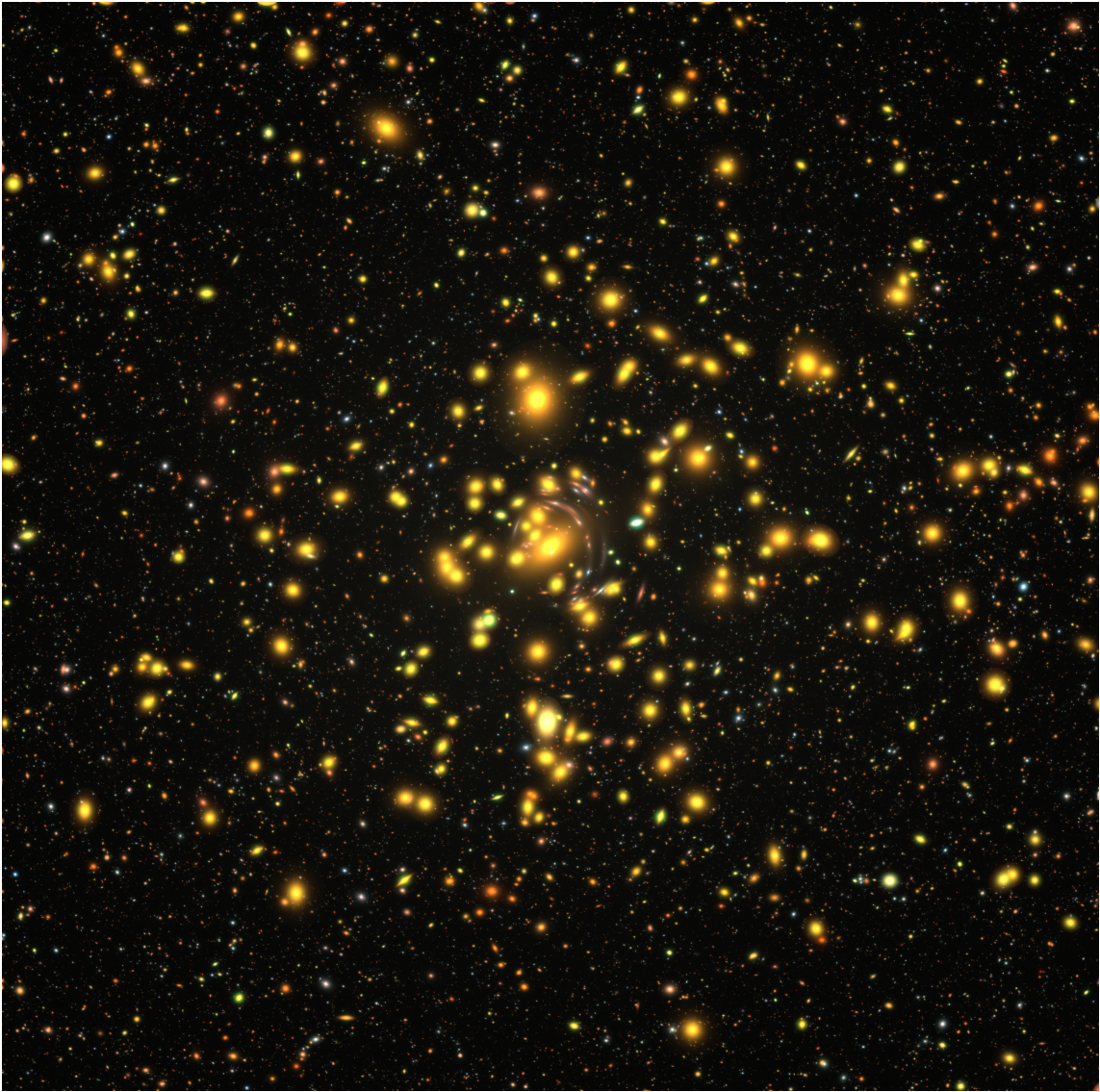}
    \caption{Superimposition of the galaxy distributions and the ICL within the full MCI FoV, presented as a composite image combining the CBU, CBV, and CBI bands. Instrumental effects, including the PSF and noise, are incorporated.}
    \label{fig-noisy}
\end{figure*}

\subsection{Instrumental Effects}

Although a detailed treatment of instrumental effects is presented in the companion paper, here we provide a simplified version of incorporating the PSF and noise into the simulated image. As detailed in Yan25, the Nyquist-sampled PSF for MCI is calculated by:
\begin{equation}
	PSF = | FFT(P)|^{2},
\end{equation}
where $P$ is the complex pupil function $P = A e^{i\cdot 2\pi OPD / \lambda}$, with the aperture function $A$, and the optical path difference ($OPD$) function at wavelength $\lambda$. This simulation for MCI PSF generates a set of 100 PSFs across the entire FoV. For simplicity, we convolve the simulated image with one of these PSFs.

The noise \citep{cao22} to be added is calculated as:
\begin{equation}
    \sigma = \frac{\sqrt{(S_{\rm img} + S_{\rm sky})\times t_{\rm expo} + N_{\rm expo}\times R^2}}{t_{\rm expo}},
\end{equation}
where $S_{\rm img}$ and $S_{\rm sky}$ are the fluxes of the sources and sky background, $N_{\rm expo}$ and $t_{\rm expo}$ are the number of exposures, and total exposure time, respectively, $R$ is the readout noise. For MCI observations, $S_{\rm sky}$ is calculated with the sky surface brightness $20.48\ \rm mag/arcsec^2$, $N_{\rm expo}=2$, $t_{\rm expo}=300\rm s$, $R=5e^{-}$.

By convolving the simulated image with the PSF and adding noise, we generate a realistic representation of the observed field (Figure~\ref{fig-noisy}). This processed image can be used by the strong lensing community to further test and refine their lens modeling codes, providing a robust tool for algorithm development and performance evaluation.

\section{Summary}\label{sect5}

We present a comprehensive simulation framework for a galaxy cluster with gravitational lensing effect for the upcoming CSST-MCI, incorporating ray-tracing for all galaxies along the line-of-sight within a wide FoV of $7.5^{\prime}\times7.5^{\prime}$. Building upon the CosmoDC2 galaxy catalog, we construct a cluster lens model consisting of main dark matter halo and subhalos corresponding to member galaxies, and we calculate the light distributions of all galaxies while accounting for lensing deflections. This simulation provides a realistic and detailed representation of a galaxy cluster deep field as observed by CSST-MCI, with primary data products including mock images in the MCI CBU, CBV, and CBI bands, the catalog of the lensing potential, and a catalog of the simulated multiple image families. The simulation framework also integrates intra-cluster light (ICL), and spectral energy distributions (SEDs) of galaxies, and instrumental effects, enhancing its applicability for further analysis. Beyond improving our understanding of the physical processes within galaxy clusters, this simulation framework serves as a crucial tool for integrating instrumental effects into the data processing pipelines. Moreover, it establishes a strong foundation for advancing cluster lens modeling and will be instrumental in future cosmological studies, particularly those related to dark matter, galaxy evolution, and large-scale structure formation using CSST-MCI data.

\normalem
\begin{acknowledgements}

The authors thank the anonymous referee who provided useful suggestions that improved this manuscript. HYS acknowledges the support from from the Ministry of Science and Technology of China (grant Nos. 2020SKA0110100), NSFC of China under grant 11973070, Key Research Program of Frontier Sciences, CAS, Grant No. ZDBS-LY-7013 and Program of Shanghai Academic/Technology Research Leader. We acknowledge the support from the science research grants from the China Manned Space Project with NO. CMS-CSST-2021-A01, CMS-CSST-2021-A04. WX thanks the support by National Key R$\&$D Program of China No. 2022YFF0503403, the support of National Nature Science Foundation of China (Nos 11988101, 12022306, 12203063), the support from the Ministry of Science and Technology of China (Nos. 2020SKA0110100), the science research grants from the China Manned Space Project (Nos. CMS-CSST-2021-B01,CMS-CSST-2021-A01), CAS Project for Young Scientists in Basic Research (No. YSBR-062), and the support from K.C.Wong Education Foundation.

\end{acknowledgements}
  
\bibliographystyle{raa}
\bibliography{bibtex}

\end{document}